# A Linear LMP Model for Active and Reactive Power with Power Loss

Yanghao Yu, Qingchun Hou, *Student Member, IEEE*, Yi Ge, Guojing Liu, and Ning Zhang, *Senior Member, IEEE*

*Abstract*— Pricing the reactive power is more necessary than ever before because of the increasing challenge of renewable energy integration on reactive power balance and voltage control. However, reactive power price is hard to be efficiently calculated because of the non-linear nature of optimal AC power flow equation. This paper proposes a linear model to calculate active and reactive power LMP simultaneously considering power loss. Firstly, a linearized AC power flow equation is proposed based on an augmented Generation Shift Distribution Factors (GSDF) matrix. Secondly, a linearized LMP model is derived using GSDF and loss factors. The formulation of LMP is further decomposed into four components: energy, congestion, voltage limitation and power loss. Finally, an iterate algorithm is proposed for calculating LMP with the proposed model. The performance of the proposed model is validated by the IEEE-118 bus system.

## Nomenclature

| | |
|---|---|
| $\mathbf{P}, \mathbf{Q}$ | Vectors of bus injected active and reactive power |
| $P_i, Q_i$ | Injected active and reactive power at bus $i$ |
| $\mathbf{Y} = \mathbf{G} + j\mathbf{B}$ | Admittance matrix of the power system |
| $\mathbf{Y}' = \mathbf{G}' + j\mathbf{B}'$ | Admittance matrix without shunt elements |
| $Y_{ij} = G_{ij} + jB_{ij}$ | Element in the $i$ th row and $j$ th column of $\mathbf{Y}$ |
| $Y'_{ij} = G'_{ij} + jB'_{ij}$ | Element in the $i$ th row and $j$ th column of $\mathbf{Y}'$ |
| $\mathbf{V}, \boldsymbol{\theta}$ | Vectors of Magnitudes and phase angles of bus voltages |
| $V_i, \theta_i$ | Voltage magnitude and phase angle at bus $i$ |
| $M, N$ | The number of branches and buses |
| $y_{ij} = g_{ij} + jb_{ij}$ | Admittance of branch $(i, j)$ |
| $y_{ii} = g_{ii} + jb_{ii}$ | Shunt admittance at bus $i$ |
| $P_m^l, Q_m^l, I_m^l$ | Power flow and current through branch $m$ |
| $P_i^G, P_i^D$ | The active power generation and active load at bus $i$ |
| $Q_i^G, Q_i^D$ | The reactive power generation and reactive load at bus $i$ |
| $F_i^P, F_i^Q$ | The fictional nodal demand at bus $i$ |
| $P_{Loss}, Q_{Loss}$ | Total power loss of the system |
| $LF_i^P, LF_i^Q$ | The active and reactive power loss factor at bus $i$ |
| $DF_i^P, DF_i^Q$ | The active and reactive power delivery factor at bus $i$ |

## I. Introduction

LOCATIONAL Marginal Price (LMP) is defined as the marginal expenses of both generators and transmission system to supply the increment of active power at each bus. A power market using LMP can provide the right price signal to relieve the congestion of power system. Therefore, LMP has been widely adopted in Independent System Operators (ISOs), such as PJM interconnection, the New York market, and the New England ISO [1].

The LMP can be divided into active power LMP (ALMP) and reactive power LMP (RLMP). The later denotes marginal cost to supply the increment of reactive power demand at each bus. Currently the LMP adopted by all of the power market is ALMP only. Most reactive power sources like static VAR compensators (SVCs) and switched capacitors (SCs) are not priced by RLMP but compensated as ancillary service in power market. Nowadays, increasing renewable energy penetration brings great challenge to power system voltage security. The adequate reactive power supply may become critical. Since the reactive power cannot be transmitted by long distance and is usually locally balanced, the need of reactive power for different node would be distinct. The value of providing reactive power would be different even there is no congestions in the power system. Therefore, introducing RLMP in the market would provide significant price signal for both short-term operation and long-term planning.

The value of LMP can be decomposed into several components. Reference [1] proposes the DCOPF model with loss to decompose LMP into the costs of marginal energy, marginal loss and congestion. Reference [2] introduces the concept of LMP to a distribution system. The distribution LMP (DLMP) is decomposed into marginal energy cost, loss cost and congestion components. References [3], [4] further analyzes both active and reactive power DLMP with consideration of congestion and voltage support. The DLMP is divided into five components: marginal costs for active power, reactive power, congestion, voltage support, and power loss. The value of different LMP components together reveal the economic and operation conditions of power system, which provide price signal to influence the behavior of participants to mitigate transmission congestion and stimulate reactive power

This work was supported by the Major Smart Grid Joint Project of National Natural Science Foundation of China and State Grid (No. U1766212) and Scientific & technical project of State Grid "Research and application of large-scale energy storage planning for receiving-end power system" (No. 5102-201918309A-0-0-00).

Y. Yu, Q. Hou, N. Zhang are with the State Key Lab of Power Systems, Department of Electrical Engineering, Tsinghua University, Beijing 100084, China (e-mail: ningzhang@tsinghua.edu.cn).
Y. Ge, G. Liu are with Economics and Technological Research Institute, State Grid Jiangsu Electric Power Co., Ltd., Nanjing, China.

supply for voltage support.

LMP can be derived from optimal power flow (OPF) [1]. In practical application of LMP, AC optimal power flow (ACOPF) and DC optimal power flow (DCOPF) are two classic models. ACOPF can simultaneously calculate ALMP and RLMP accurately. But it is a non-convex problem which may be computationally intractable even for smaller systems [5]. Additionally, LMP derived from ACOPF are not analytical so that it cannot be decomposed into different components and is lack of transparency [6]. Comparatively, DCOPF model is more concise and efficient, which makes it widely utilized by LMP-based markets. However, since it neglects the voltage fluctuation and power loss, it has a relatively low accuracy in LMP calculation. In addition, DCOPF cannot calculate RLMP since it ignores the reactive power balance and reactive power flow.

Recent researches have been conducted to improve the LMP calculation. Reference [7] uses energy reference bus independent model to obtain a more accurate calculation on energy, congestion and loss components of LMP. In Reference [8], an iterative DCOPF algorithm is proposed to address the nonlinear marginal loss. Reference [9] proposes a linear loss-embedded model to calculate LMP with loss components by dealing with marginal generators under lossless DCOPF solution. Reference [10] proposes a linear-constrained and convergence-guaranteed OPF model that explicitly considers the constraints of reactive power balance and bus voltage. Reference [11] proposes a linearized power flow model with decoupling of voltage magnitude and phase angle to achieve accurate estimation of voltage magnitude. Based on the linearized power flow model, Reference [6] further proposes the decoupled OPF model using Generation Shift Distribution Factors (GSDF) and further applies it to jointly calculate ALMP and RLMP with respect to voltage constraints, but the impact of network loss is neglected in this model.

This paper proposes a linearized iterative model to calculate both ALMP and RLMP considering power loss. A linearized AC power flow equation is proposed based on the augmented GSDF matrix. The active and reactive power loss are linearized by introducing loss factor (LF). An iterate algorithm is proposed to calculate power loss and the LMP. Finally, the accuracy and performance of the proposed model is validated on an IEEE-118 bus case.

The rest of this paper is organized as follows: Section II constructs the linearized LMP model. Section III validates the accuracy of the ALMP and RLMP calculation of the model through the IEEE 118-bus case on MATPOWER. Finally, Section IV concludes the paper.

## II. LINEARIZED LMP MODEL WITH POWER LOSS

This Section builds the LMP model based on GSDF. The modeling is divided into five subproblems: 1) Introducing GSDF-based linear AC power flow model; 2) Linearizing the power losses; 3) Building GSDF-based LMP model; 4) Deriving ALMP and RLMP based on the LMP model; 5) Proposing iterate algorithm to solve LMP. This Section will present the technical details of each subproblem.

### A. Power Flow Equations with GSDF

A linearized power flow equation with respect to voltage magnitude and phase angle is proposed by reference [11] as follows.

$$\begin{bmatrix} \mathbf{P} \\ \mathbf{Q} \end{bmatrix} = -\begin{bmatrix} \mathbf{B}' & -\mathbf{G} \\ \mathbf{G}' & \mathbf{B} \end{bmatrix} \begin{bmatrix} \boldsymbol{\theta} \\ \mathbf{V} \end{bmatrix} \triangleq -\mathbf{C} \begin{bmatrix} \boldsymbol{\theta} \\ \mathbf{V} \end{bmatrix} \quad (1)$$

The matrix $\mathbf{C}$ in Eq. (1) is a $2N \times 2N$ matrix determined solely by the element of admittance matrix.

Based on Eq. (1), the active power balance equation is derived as follows:

$$\sum_{i=1}^{N} P_i = \sum_{i=1}^{N}\sum_{j=1}^{N} G_{ij}V_j - \sum_{i=1}^{N}\sum_{j=1}^{N} B'_{ij}\theta_j = \sum_{j=1}^{N}\left(\sum_{i=1}^{N} G_{ij}\right)V_j - \sum_{j=1}^{N}\left(\sum_{i=1}^{N} B'_{ij}\right)\theta_j$$
$$= \sum_{j=1}^{N} g_{jj}V_j \approx \sum_{j=1}^{N} g_{jj} \approx 0 \quad (2)$$

The two approximation steps in Eq. (2) are based on the assumptions that $V_j \approx 1$ and $g_{jj} \approx 0$. Similarly, we have the reactive power balance equation as follows:

$$\sum_{i=1}^{N} Q_i = -\sum_{j=1}^{N} b_{jj}V_j \approx -\sum_{j=1}^{N} b_{jj} \quad (3)$$

According to Eq. (2), the equations Eq. (1) is not independent. In other words, matrix $\mathbf{C}$ in (1) is close to singular. So we remove the row and column of the reference bus in $\mathbf{C}$ and inverse the new $(2N-1) \times (2N-1)$ matrix. We then fill the row and column of the reference bus in the inversed matrix with zeros. Hence, we have a new $2N \times 2N$ matrix $\mathbf{X}$ satisfying:

$$\begin{bmatrix} \boldsymbol{\theta} \\ \mathbf{V} \end{bmatrix} = \mathbf{X} \begin{bmatrix} \mathbf{P} \\ \mathbf{Q} \end{bmatrix} \quad (4)$$

Unfolding the matrix form of Eq. (4), the voltage equation is acquired as follows:

$$V_i = \sum_{k=1}^{N} X_{N+i,k} P_k + \sum_{k=1}^{N} X_{N+i,N+k} Q_k \quad (5)$$

According to Eq. (4), the power flow of branches can be linearly expressed by the bus power injection both of active and reactive parts:

$$P_m^l = P_{ij} \approx (V_i - V_j)g_{ij} - b_{ij}(\theta_i - \theta_j)$$
$$= \sum_{k=1}^{N} \left(g_{ij}(X_{N+i,k} - X_{N+j,k}) - b_{ij}(X_{i,k} - X_{j,k})\right)P_k \quad (6)$$
$$+ \sum_{k=1}^{N} \left(g_{ij}(X_{N+i,N+k} - X_{N+j,N+k}) - b_{ij}(X_{i,N+k} - X_{j,N+k})\right)Q_k$$

$$Q_m^l = Q_{ij} \approx -b_{ij}(V_i - V_j) - g_{ij}(\theta_i - \theta_j)$$
$$= \sum_{k=1}^{N} \left(-g_{ij}(X_{i,k} - X_{j,k}) - b_{ij}(X_{N+i,k} - X_{N+j,k})\right)P_k \quad (7)$$
$$+ \sum_{k=1}^{N} \left(-g_{ij}(X_{i,N+k} - X_{j,N+k}) - b_{ij}(X_{N+i,N+k} - X_{N+j,N+k})\right)Q_k$$

The coefficients of $P_k$ and $Q_k$ in Eqs. (6)-(7) forms the GSDF matrix between the branch power flow and power injection:

$$GSF_{m-k}^{P-P} = g_{ij}\left(X_{N+i,k} - X_{N+j,k}\right) - b_{ij}\left(X_{i,k} - X_{j,k}\right)$$
$$GSF_{m-k}^{P-Q} = g_{ij}\left(X_{N+i,N+k} - X_{N+j,N+k}\right) - b_{ij}\left(X_{i,N+k} - X_{j,N+k}\right)$$
$$GSF_{m-k}^{Q-P} = -g_{ij}\left(X_{i,k} - X_{j,k}\right) - b_{ij}\left(X_{N+i,k} - X_{N+j,k}\right)$$
$$GSF_{m-k}^{Q-Q} = -g_{ij}\left(X_{i,N+k} - X_{j,N+k}\right) - b_{ij}\left(X_{N+i,N+k} - X_{N+j,N+k}\right)$$
(8)

Finally, the active and reactive power flow equations (Eqs. (6)-(7)) can be expressed using GSDF:

$$P_m^l = \sum_{i=1}^{N} GSF_{m-i}^{P-P} P_i + \sum_{i=1}^{N} GSF_{m-i}^{P-Q} Q_i$$
$$= \sum_{i=1}^{N} GSF_{m-i}^{P-P}\left(P_i^G - P_i^D\right) + \sum_{i=1}^{N} GSF_{m-i}^{P-Q}\left(Q_i^G - Q_i^D\right), \text{ for } m = 1, 2 \cdots M$$
$$Q_m^l = \sum_{i=1}^{N} GSF_{m-i}^{Q-P} P_i + \sum_{i=1}^{N} GSF_{m-i}^{Q-Q} Q_i$$
$$= \sum_{i=1}^{N} GSF_{m-i}^{Q-P}\left(P_i^G - P_i^D\right) + \sum_{i=1}^{N} GSF_{m-i}^{Q-Q}\left(Q_i^G - Q_i^D\right), \text{ for } m = 1, 2 \cdots M$$
(9)

Different from classic DC power flow model, the proposed model considers both active and reactive power, which provide foundations for deriving ALMP and RLMP.

*B. Considering Power Loss in GSDF-based Power Flow Equations*

It should be noted that in the derivation of last subsection we do not consider power losses. In other words, the total active power generation and reactive power generation in Eq. (2) equals the active load and reactive load respectively (plus the load on the shunt branch). Such lossless consumption does not coincide with practical power system. Therefore, we need to consider the power losses in power balance equations and allocate the power losses to each node as virtual load.

According to Joule's Law, both active and reactive power losses, can be calculated using branch power flow (assuming that $V_i \approx 1$):

$$P_{Loss} = \sum_{m=1}^{M}\left(I_m^l\right)^2 R_m \approx \sum_{m=1}^{M}\left(P_m^l\right)^2 R_m$$
$$= \sum_{m=1}^{M}\left(\sum_{i=1}^{N} GSF_{m-i}^{P-P}\left(P_i^G - P_i^D\right) + \sum_{i=1}^{N} GSF_{m-i}^{P-Q}\left(Q_i^G - Q_i^D\right)\right)^2 R_m$$
$$Q_{Loss} = \sum_{m=1}^{M}\left(I_m^l\right)^2 X_m \approx \sum_{m=1}^{M}\left(P_m^l\right)^2 X_m$$
$$= \sum_{m=1}^{M}\left(\sum_{i=1}^{N} GSF_{m-i}^{P-P}\left(P_i^G - P_i^D\right) + \sum_{i=1}^{N} GSF_{m-i}^{P-Q}\left(Q_i^G - Q_i^D\right)\right)^2 X_m$$
(10)

In order to consider power losses, the concept of loss factor (LF), delivery factor (DF) and fictional nodal demand (FND) [8] are introduced in this paper. LF indicates the marginal power losses with respect to an incremental power injection on a bus, while DF indicate the active and reactive power that is equivalently injected to the grid. LF and DF for active and reactive power can be stated as Eq. (11).

$$LF_i^P = \frac{\partial P_{Loss}}{\partial P_i^G} = \sum_{m=1}^{M} 2P_m^l GSF_{m-i}^{P-P} R_m, \ DF_i^P = 1 - LF_i^P \text{ for } i = 1, 2 \cdots N$$
$$LF_i^Q = \frac{\partial Q_{Loss}}{\partial Q_i^G} = \sum_{m=1}^{M} 2P_m^l GSF_{m-i}^{P-Q} X_m, \ DF_i^Q = 1 - LF_i^Q \text{ for } i = 1, 2 \cdots N$$
(11)

It should be noted that $LF_i^P$ and $LF_i^Q$ may be negative for some node, denoting that the power injection would reduce the power losses. We introduce the DF into the power balance Eqs. (2)-(3) so that the power losses can be explicitly formulated in the equation:

$$\sum_{i=1}^{N} DF_i^P\left(P_i^G - P_i^D\right) + P_{Loss} = 0$$
$$\sum_{i=1}^{N} DF_i^Q\left(Q_i^G - Q_i^D\right) + Q_{Loss} = -\sum_{j=1}^{N} b_{jj}$$
(12)

FND is introduced to evenly allocate the loss of each branch to its two end buses, mathematically:

$$F_i^P = \frac{1}{2}\sum_{i \cap m \neq \varnothing}\left(P_m^l\right)^2 R_m, F_i^Q = \frac{1}{2}\sum_{i \cap m \neq \varnothing}\left(P_m^l\right)^2 X_m, \text{ for } i = 1, 2 \cdots N \quad (13)$$

Each bus takes the $F_i^P$ and $F_i^Q$ on their connected branches as a virtual injection. Therefore, the power injection at each bus is restated as:

$$P_i = P_i^G - P_i^D - F_i^P \text{ for } i = 1, 2 \cdots N$$
$$Q_i = Q_i^G - Q_i^D - F_i^Q \text{ for } i = 1, 2 \cdots N$$
(14)

By introducing the FND in each bus, the total active power generation no longer equals to the active load since the generation need to balance the extra virtual injection. It suggests that both the active and reactive power losses are considered in the proposed GSDF-based power flow equations.

*C. LMP Model Based on GSDF*

According to the derivations above, especially with Eqs. (5), (9), (12), and (14), we draw the GSDF-based LMP model as shown in Eq. (15). $a_i, b_i, c_i$ represent the cost coefficients of generator at bus $i$, the cost functions of active and reactive power generation are both quadratic while reactive function does not have primary term; $\lambda_P, \lambda_Q, u_k, v_i$ denote the Lagrange multiplier of the constraints of active power balance, reactive power balance, power flow congestion and voltage respectively; $DF_i^P, DF_i^Q$ denote the DF at bus $i$; $P_{Loss}, Q_{Loss}$ are the total system active and reactive power loss; $F_i^P, F_i^Q$ are active and reactive FND at bus $i$; $P_k^{\lim}$ is the active power limits of branch $k$; $V_i^{\min}, V_i^{\max}$ denote the minimum and maximum voltage at bus $i$ respectively; $P_i^{G\min}, P_i^{G\max}, Q_i^{G\min}, Q_i^{G\max}$ denote the active and reactive power generation limits at bus $i$ respectively.

$$Min \ \sum_{i=1}^{N} a_i P_i^G + b_i \left(P_i^G\right)^2 + c_i \left(Q_i^G\right)^2$$

$$s.t. \ f_P \triangleq \sum_{i=1}^{N} DF_i^P \left(P_i^G - P_i^D\right) + P_{Loss} = 0 \quad \lambda_P$$

$$f_Q \triangleq \sum_{i=1}^{N} DF_i^Q \left(Q_i^G - Q_i^D\right) + Q_{Loss} = -\sum_{j=1}^{N} b_{jj} \quad \lambda_Q$$

$$g_m \triangleq \sum_{i=1}^{N} GSF_{m-i}^{P-P} \left(P_i^G - P_i^D + F_i^P\right) + \sum_{i=1}^{N} GSF_{m-i}^{P-Q} \left(Q_i^G - Q_i^D + F_i^Q\right) = P_m^l, \text{ for } m = 1, 2 \cdots M \quad u_m$$

$$h_i^V \triangleq \sum_{k=1}^{N} X_{N+i,k} \left(P_k^G - P_k^D + F_k^P\right) + \sum_{k=1}^{N} X_{N+i,N+k} \left(Q_k^G - Q_k^D + F_k^Q\right) = V_i, \text{ for } i = 1, 2 \cdots N \quad v_i \quad (15)$$

$$-P_m^{\lim} \leq P_m^l \leq P_m^{\lim}, \text{ for } m = 1, 2 \cdots M$$

$$V_i^{\min} \leq V_i \leq V_i^{\max}, \text{ for } i = 1, 2 \cdots N$$

$$P_i^{G\min} \leq P_i^G \leq P_i^{G\max}, \text{ for } i = 1, 2 \cdots N$$

$$Q_i^{G\min} \leq Q_i^G \leq Q_i^{G\max}, \text{ for } i = 1, 2 \cdots N$$

*D. ALMP and RLMP Decomposition*

When the optimal solution of the model is reached, LMP can be calculated by the Lagrange Function:

$$\psi = \left(\sum_{i=1}^{N} a_i P_i^G + b_i \left(P_i^G\right)^2 + c_i \left(Q_i^G\right)^2\right) - \lambda_P f_P - \lambda_Q f_Q - \sum_{m=1}^{M} \mu_m g_m - \sum_{i=1}^{N} v_i h_i^V \quad (16)$$

ALMP is defined as the marginal cost of an incremental change of active power demand at a bus. ALMP at bus $i$ can be calculated as the first-order partial derivative of the Lagrange Function with respect to the active power load at bus $i$:

$$ALMP_i = \frac{\partial \psi}{\partial P_i^D}$$
$$= -\lambda_P \frac{\partial f_P}{\partial P_i^D} - \sum_{m=1}^{M} \mu_m \frac{\partial g_m}{\partial P_i^D} - \sum_{j=1}^{N} v_j \frac{\partial h_j^V}{\partial P_i^D} \quad (17)$$
$$= DF_i^P \lambda_P + \sum_{m=1}^{M} \mu_m GSF_{m-i}^{P-P} + \sum_{j=1}^{N} v_j X_{N+j,i}$$
$$= \lambda_P + \sum_{m=1}^{M} \mu_m GSF_{m-i}^{P-P} + \sum_{j=1}^{N} v_j X_{N+j,i} - LF_i^P \lambda_P$$

Similarly, RLMP is defined as marginal cost of an incremental change of reactive power demand at a bus. The calculation of RLMP at bus $i$ is:

$$RLMP_i = \frac{\partial \psi}{\partial Q_i^D}$$
$$= -\lambda_Q \frac{\partial f_Q}{\partial Q_i^D} - \sum_{m=1}^{M} \mu_m \frac{\partial g_m}{\partial Q_i^D} - \sum_{j=1}^{N} v_j \frac{\partial h_j^V}{\partial Q_i^D} \quad (18)$$
$$= DF_i^Q \lambda_Q + \sum_{m=1}^{M} \mu_m GSF_{m-i}^{P-Q} + \sum_{j=1}^{N} v_j X_{N+j,N+i}$$
$$= \lambda_Q + \sum_{m=1}^{M} \mu_m GSF_{m-i}^{P-Q} + \sum_{j=1}^{N} v_j X_{N+j,N+i} - LF_i^Q \lambda_Q$$

Eqs (17) and (18) derives the analytical form different components of ALMP and RLMP. The first term of LMP is the energy component derived from the power balance constraints. The second term is the congestion component which is related to the power flow equations. The third term is the voltage constraint component. The last term is loss component. If loss is not considered in a model, LF is zero. When the constraints on power flow or voltage magnitudes are active, their corresponding components of LMP are non-zero.

*E. Solving Algorithm*

An iteration algorithm is proposed to solve the LMP model with power loss, as shown in Fig. 1. First, we solve the LMP model without power loss. Then the power loss is calculated using Eqs. (10)-(11) and is allocated using Eqs. (14). The LMP model is later updated with the calculated DF and FND. After one iteration, we solve the updated LMP model and re-calculate the power losses. When the difference of power losses in two iterations is smaller than a certain threshold, iteration stops and a final LMP is obtained from the newest LMP model.

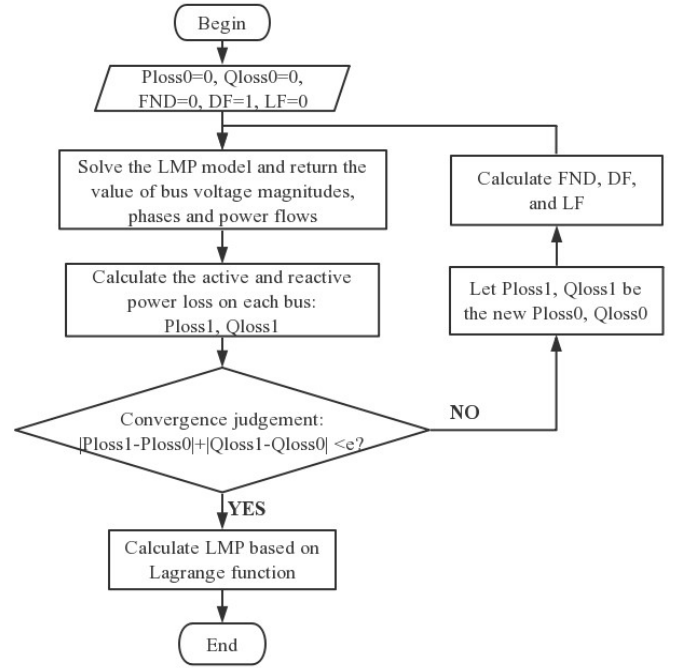

Figure 1 Flowchart of the iteration algorithm

III. CASE STUDY

In this section, the GSDF-based LMP model is applied to an IEEE 118-bus case on MATPOWER 6.0 to validate its accuracy in calculating ALMP and RLMP. The model is implemented in Matlab R2016a with the solver Gurobi8.0. The simulation platform is a personal computer with an Intel Core i7 CPU @2.50GHz and 8GB RAM.

The IEEE 118-bus test case is from MATPOWER 6.0 toolbox of MATLAB [12]. Meanwhile, the original case does not contain limits on branch power flow. The branch MVA rating data are taken from Index of Data Illinois Institute of Technology [13]. Moreover, the case does not provide cost for reactive power. We add the reactive cost function of generators and let $c_i$ in Eq. (15) be the same as $b_i$.

The LMP results of our model (denoted as LMP-L) are compared with the LMP model without loss (denoted as LMP-O), and the LMP results derived from ACOPF and DCOPF solver in MATPOWER (denoted as LMP-AC and LMP-DC). The result obtained from ACOPF is set as the benchmark result to calculate the accuracy of other results.

### A. ALMP Accuracy Evaluation.

Since the ALMP is mostly concerned in the market, we compare the accuracy of ALMP obtained from the proposed method and that from result derived from LMP-O and DCOPF, taking ACOPF as the standard result. The index of the average error of ALMP (AEA) is introduced as:

$$AEA = \frac{\sum_{i=1}^{N} \left| \left( ALMP_i - ALMP_i^{AC} \right) / ALMP_i^{AC} \right|}{N} \quad (19)$$

Here $ALMP_i^{AC}$ is the ALMP result at bus $i$ from ACOPF.

In the ACOPF model, when the voltage constraint is active, the voltage component influences the LMP. In order to demonstrate the effectiveness of our proposed model in considering the voltage constraint, three scenarios with different voltage constraints are carried out in the case study:
1) Loose voltage constraint: $V_{min} = 0.90, V_{max} = 1.10$;
2) Normal voltage constraint: $V_{min} = 0.95, V_{max} = 1.05$;
3) Tight voltage constraint: $V_{min} = 0.97, V_{max} = 1.03$.

Fig.2 shows the AEA of the loss-embedded LMP model (ALMP-L), LMP model without loss (ALMP-O) and DCOPF model (ALMP-DC) when load level changes in different voltage constraint scenarios. ALMP-O performs almost the same as ALMP-DC in loose and normal scenarios and slightly better than ALMP-DC in tight scenario. This indicates that the linearized model without power loss is comparable with DCOPF. In addition, the ALMP-L results outperform both ALMP-O and ALMP-DC in all scenarios and under all load levels. At 0.95 p.u. load level, the AEA of ALMP-DC is above 3% in all scenarios, while the AEA of ALMP-L is around 1.5%. LOPF performs 50% better than DCOPF. As load level rises, AEA of ALMP-L has a trend of increasing, but it still outperforms DCOPF on ALMP. The results show that the proposed linearize LMP model with power loss shows better accuracy in calculating ALMP when comparing with DCOPF, and power loss component plays an important role in ALMP accuracy.

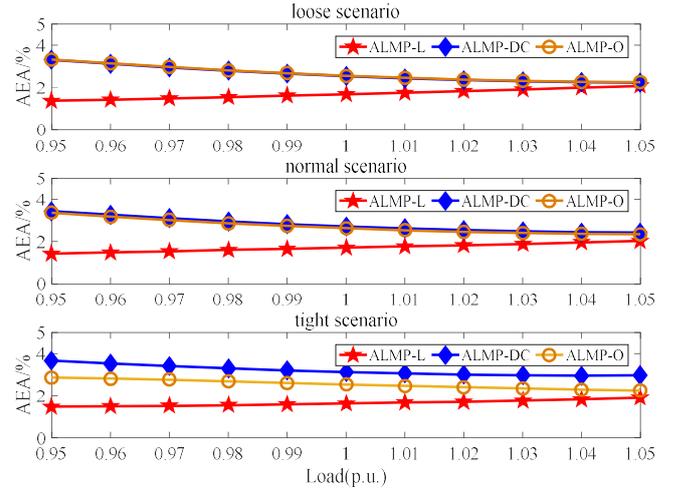

Figure 2 AEA in three voltage scenarios

### B. ALMP and RLMP Analysis

The case in tight voltage constraint is further analyzed. Fig.3 presents the ALMP and RLMP at each bus from LMP-L, DCOPF and ACOPF respectively (DCOPF does not have RLMP). LMP-L tracks the fluctuation of ALMP derived from ACOPF at most buses. However, DCOPF gives a fixed ALMP value at all buses, thus it has massive error compared with ACOPF. One of the reasons is that DCOPF does not consider the limit of bus voltage and reactive power. In the meanwhile, the RLMP results from LMP-L model is almost the same as that from ACOPF, which indicates that model also performs well in RLMP calculation.

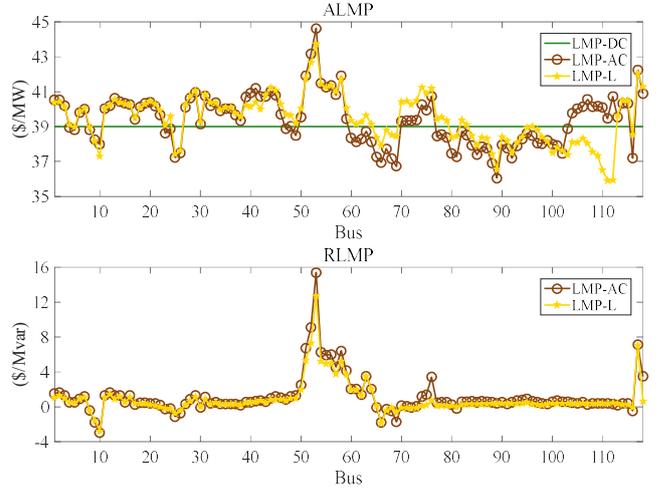

Figure 3 ALMP and RLMP in tight voltage constraint

Fig. 4 illustrates that voltage magnitudes calculated from LMP-L are also accurate comparing to LMP-AC. Moreover, it indicates whether the voltage magnitude at certain bus reaches its boundary or not, which is clearly reflected on the LMP value as voltage component. For example, the voltage magnitudes at bus 10, 25, 66 and 69 reach their upper limit, which means they already have abundant reactive power. Consequently, RLMP in Fig. 3 are relatively low at these buses. On the contrary, the voltage magnitudes at bus 58 and 117 reach their lower limit, making their RLMP relatively higher. In fact, ALMP follows the same rule as RLMP but it's not as apparent as RLMP shown in Fig. 3. As a result, LMP especially RLMP can work as an indicator of reactive power demand. It further provides incentives for the reactive power compensators to join the market.

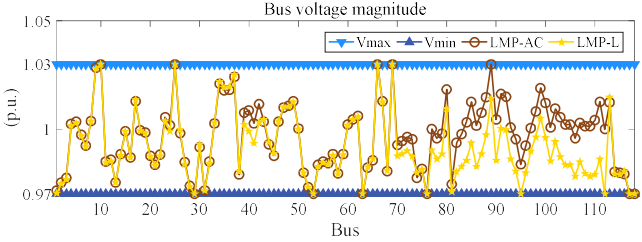

Figure 4 Bus voltage magnitude at each bus

Moreover, Fig. 5 shows the active and reactive power generated by all generators. The active power generation result of LMP-L is closer to ACOPF than DCOPF. As for the reactive power generation, LMP-L is nearly the same as that from ACOPF at the marginal generators such as generator NO. 23, 24, and 35. The marginal generators altogether determine the energy term of LMP which is the majority of LMP. Therefore, locating the right marginal generators greatly improves the LMP results.

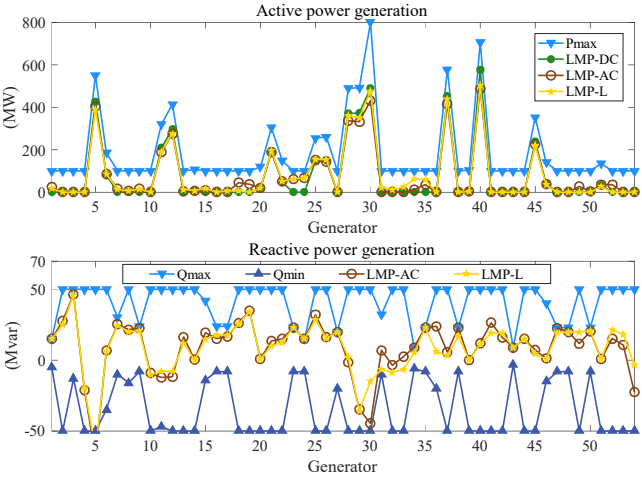

Figure 5 Generators output

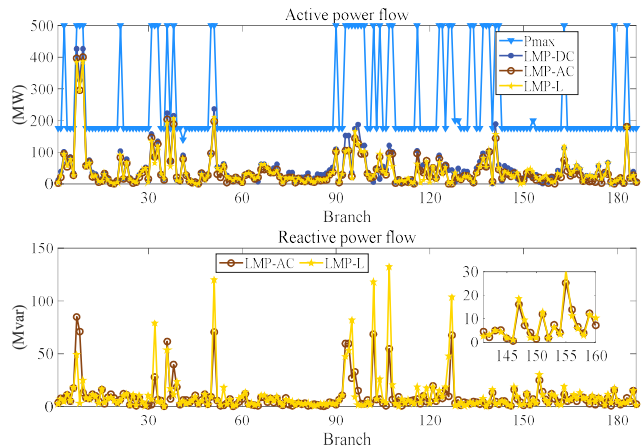

Figure 6 Active and reactive power flows

Fig. 6 presents the active and reactive power flow through all 186 branches. It indicates that none of the branch limits is reached. Thus, the congestion component does not have contribution to LMP in this case. That's another reason why the DCOPF gives a fixed ALMP value at all buses.

Above all, through the case study on IEEE-118 bus system, this model is proved to be more accurate than DCOPF when calculating ALMP. This model also shows high accuracy in RLMP calculation compared with ACOPF model.

## IV. CONCLUSION

In this paper, a GSDF-based LMP model for both ALMP and RLMP calculation with power loss is proposed. The necessity to price the reactive power is first discussed. Then a linearized LMP model is derived from an augmented GSDF matrix. Loss equations and loss-related factors are added to enrich the model. Through derivation of Lagrange function, LMP can be decomposed to four components (energy, congestion, voltage limitation and power loss). A case study on the IEEE-118 bus case validates the accuracy of our model in LMP calculation. The case demonstrates that ACOPF constraints can be clearly reflected on LMP components. Therefore, with the congestion and voltage components of LMP, the operator can effectively manage the branch congestion and voltage support respectively. It is also indicated that the power loss component in our linear LMP model is effective to improve the LMP results.